\documentclass[aps,prb,twocolumn,superscriptaddress,groupedaddress,showpacs]{revtex4}  
\usepackage{amsmath}
\usepackage{amsfonts}
\usepackage{amssymb}
\usepackage{graphics}
\usepackage{graphicx}
\usepackage{color}
\usepackage[cyr]{aeguill}

\newcommand{\ii}{\text{i}}

\definecolor{green2}{rgb}{0,0.66,0}

\begin{document}

\title{Acoustic double negativity induced by position correlations within a disordered set of monopolar resonators}
\author{Maxime Lanoy}
\affiliation{Department of Physics and Astronomy, University of Manitoba, Winnipeg, Manitoba R3T 2N2, Canada}
\affiliation{Institut Langevin, ESPCI ParisTech, CNRS (UMR 7587), PSL Research University, Paris, France}
\affiliation{Laboratoire Mati\`ere et Syst\`emes Complexes, Universit\'e Paris-Diderot, CNRS (UMR 7057), Paris, France}

\author{John H. Page}
\affiliation{Department of Physics and Astronomy, University of Manitoba, Winnipeg, Manitoba R3T 2N2, Canada}

\author{Geoffroy Lerosey}
\affiliation{Institut Langevin, ESPCI ParisTech, CNRS (UMR 7587), PSL Research University, Paris, France}

\author{Fabrice Lemoult}
\affiliation{Institut Langevin, ESPCI ParisTech, CNRS (UMR 7587), PSL Research University, Paris, France}

\author{Arnaud Tourin}
\affiliation{Institut Langevin, ESPCI ParisTech, CNRS (UMR 7587), PSL Research University, Paris, France}

\author{Valentin Leroy}
\affiliation{Laboratoire Mati\`ere et Syst\`emes Complexes, Universit\'e Paris-Diderot, CNRS (UMR 7057), Paris, France}


\date{\today}

\begin{abstract}
Using a Multiple Scattering Theory algorithm, we investigate numerically the transmission of ultrasonic waves through a disordered locally resonant metamaterial containing only monopolar resonators. By comparing the cases of a perfectly random medium with its pair correlated counterpart, we show that the introduction of short range correlation can substantially impact the effective parameters of the sample. We report, notably, the opening of an acoustic transparency window in the region of the hybridization band gap. Interestingly, the transparency window is found to be associated with negative values of both effective compressibility and density. Despite this feature being unexpected for a disordered medium of monopolar resonators, we show that it can be fully described analytically and that it gives rise to negative refraction of waves.
\end{abstract}


\maketitle

The interaction between waves and matter is increasingly exploited, whether it be for shaping wavefields~\cite{vellekoop2007focusing}, focusing~\cite{cassereau1992time}, absorbing~\cite{tao2008metamaterial,landy2008perfect, mei2012dark} or cloaking~\cite{alu2005achieving,leonhardt2006optical,pendry2006controlling}. Among all these challenges, there continues to be extensive interest in a promising route for superresolution focusing via the design and realization of left-handed (or doubly negative) materials, as proposed by Pendry~\cite{pendry2000negative}. Such behavior has been realized through the development of locally resonant metamaterials that consist of an assembly of subwavelength resonant inclusions inside a continuous matrix~\cite{shelby2001experimental, smith2000composite}.  Exploiting the band folding inside a periodic structure~\cite{notomi2000theory,mizuguchi2003focusing,foteinopoulou2003refraction,cubukcu2003electromagnetic,parimi2003photonic,yang2004focusing,zhang2004negative,sukhovich2008negative,sukhovich2009experimental} turned out to be a rewarding alternative strategy to observe high resolution focusing.
By contrast, locally resonant metamaterials have the advantage that their subwavelength internal structure enables their description by effective medium theories. Because the resonances occur at very low frequencies, the effective properties of metamaterials are usually believed to rely on the individual features of the scatterers and on their concentration rather than on their spatial distribution.
As a result, opportunities for tailoring their properties by exploiting both subwavelength resonators \emph{and} the way they are spatially distributed remain largely unexplored.

For acoustic waves, the relevant parameters to describe the effective properties are the density $\rho_\text{eff}$ and compressibility $\chi_\text{eff}$.
By going beyond the isotropic scatterer approximation employed in Foldy's~\cite{foldy} seminal work, Waterman and Truell~\cite{waterman1961multiple} proposed a relationship that only depends on the concentration $n$, and on the forward $f(0)$ and backward $f(\pi)$ scattering functions of the inclusions:
\begin{subequations}
\label{eqWT}
\begin{align}
\frac{\chi_\text{eff}}{\chi_0} & =  1+\frac{2 \pi n }{k_0^2}\big[ f(0)+f(\pi) \big],\\
\frac{\rho_\text{eff}}{\rho_0} & =  1+\frac{2 \pi n }{k_0^2}\big[ f(0)-f(\pi)\big],
\label{rhochi}
\end{align}
\end{subequations}
where $\chi_0$, $\rho_0$ and $k_0$ are the compressibility, the density and the wavenumber in the host medium. Equation~(\ref{eqWT}a) indicates that negative compressibility is relatively easy to obtain. Indeed, for scatterers exhibiting a monopolar resonance, such as gas bubbles, $\text{Re}[f(0)+f(\pi)]$ can reach large negative values so that $\text{Re}[\chi_\text{eff}]<0$. Monopolar scattering, however, will not affect the effective density because it gives $f(0)-f(\pi)=0$. Thus, to achieve double negativity, it is necessary to invoke higher mode resonances such as dipolar ones. A number of previous studies~\cite{ding2007metamaterial,lee2010composite} have shown that incorporating two different resonators, one monopolar and the other dipolar, within the unit cell of a metamaterial can be a successful strategy. Equally convincing results were obtained using the original idea of designing a single meta-inclusion that featured both kind of resonances for overlapping frequency ranges~\cite{li2004double,brunet2015soft}. However, the last requirement can be particularly challenging for acoustic waves and reduces the possibilities to a small handful of inclusion types. Moreover, the dipolar mode is not easy to excite in the long wavelength regime meaning that one is limited in terms of miniaturization. 

In this article, we introduce another strategy without the requirement that the individual inclusions have both monopolar and dipolar resonances, and demonstrate how a three-dimensionnal (3D) doubly negative metamaterial can be created that is populated \emph{solely} with monopolar subwavelength resonators. Starting with a random distribution of these resonators, we impose a pair-wise spatial correlation between them. We show, both numerically and analytically, that multiple scattering coupling within each pair leads to the creation of two modes: a symmetrical mode, which influences the compressibility, and an antisymmetric mode, which influences the dynamic mass density. Thus, these pair-wise correlations enable the conditions needed for acoustic double negativity to be achieved. To show these effects convincingly,  we compare a perfectly random sample, for which the interferences between the incident and the scattered fields lead to the opening of a band gap above the resonance frequency, with a pair-correlated collection of the same scatterers. We demonstrate that such pairing significantly impacts the propagation and is responsible for the appearance of a transparency window within the band gap. We carefully study the effective parameters of both kinds of samples and present evidence that the spatial correlations between scatterer pairs lead to simultaneously negative values of the dynamic mass density $\rho_\text{eff}$ and compressibility $\chi_\text{eff}$ inside this transparency window; note that this occurs even though negative values of $\rho_\text{eff}$ are usually acknowledged to be unreachable for monopolar materials. We verify that an acoustic wave, with a frequency lying inside the transparency window, is negatively refracted as it impiges on a slab of randomly distributed paired scatterers. Finally we discuss the influence of dissipation on the doubly negative behavior and apply our results to a suspension of bubbles in a yield stress fluid.

We consider a monopolar acoustic scatterer characterized by its radius $a$ and an isotropic scattering function given by the Lorentzian law:
\begin{equation}
f(\omega)=\frac{- a}{1-\omega_0^2/\omega^2+\text{i}(k_0 a  + \delta)},
\label{lor}
\end{equation}
with $\omega_0$ being the monopolar resonance frequency and $\delta$ the damping factor accounting for dissipation. We first investigate the case of a conservative scattering process for which $\delta=0$~\cite{mie1908beitrage}.
For a sample consisting of $N$ scatterers for which the positions and radii are known, one can apply the Multiple Scattering Theory (MST) as introduced by Lax~\cite{lax1952multiple}.
Because the calculation turns out to become very demanding in the case of a high number of inclusions, we have developed a numerical algorithm which was introduced in Ref.~\onlinecite{lanoy2015subwavelength} and enables us to fully determine the acoustic linear response to any kind of excitation for a sample containing up to $30,000$ scatterers.

\begin{figure}[htb!]
    \centering
      \includegraphics[width=\linewidth]{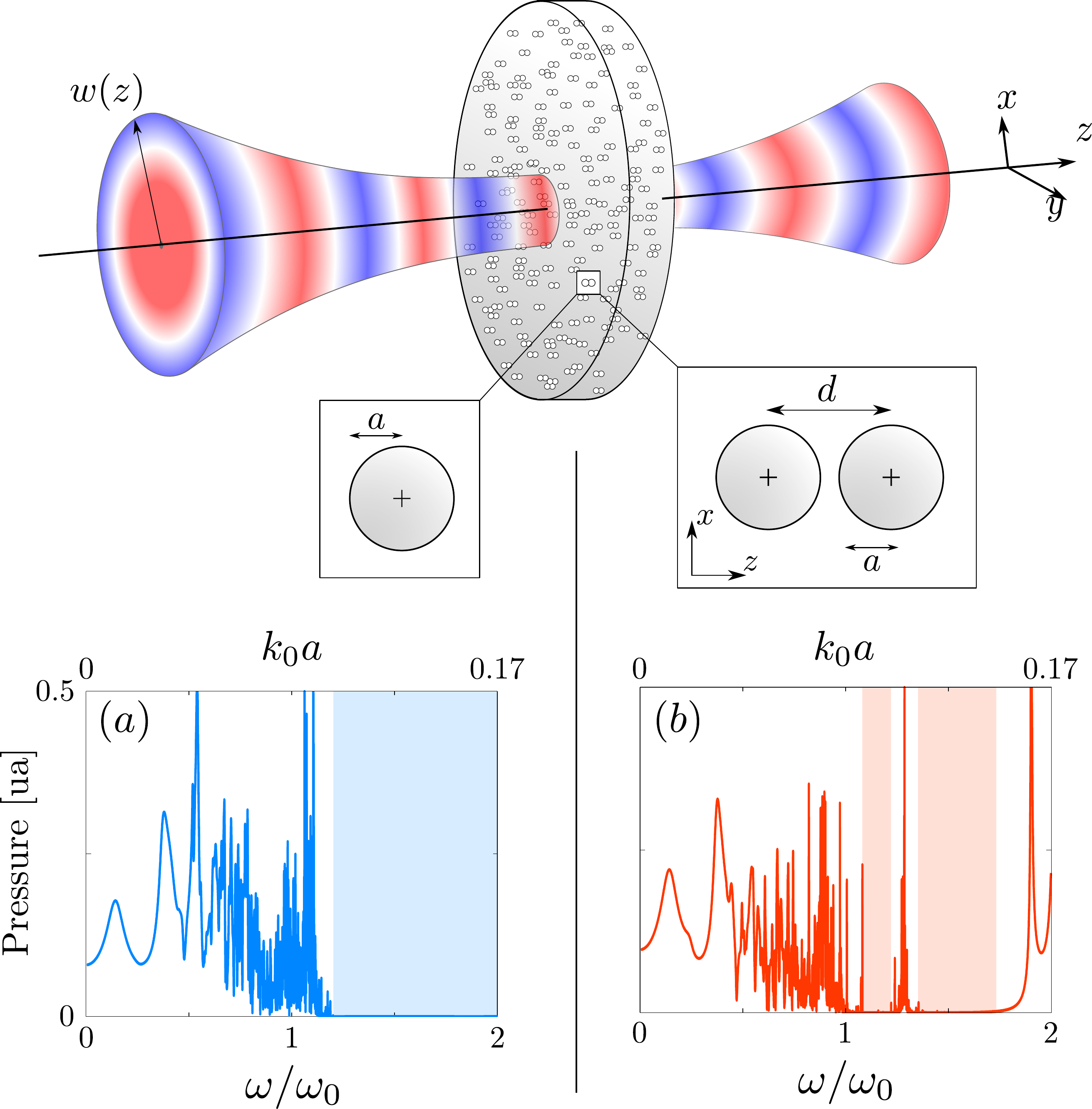}
    \caption{Schematic illustration of the numerical setup together with the pressure field spectra obtained for two different configurations: (a) a spatially random distribution of non-interpenetrating scatterers, and (b) a spatially random distribution of pairs of scatterers separated by a distance $d=250~\mu$m and aligned parallel to the z-axis. In both cases, $a=100~\mu$m and $n=2.38~\text{mm}^ {-3}$.}\label{spectre}
\end{figure}
From now on, we consider the following set of parameters: $a=100~\mu$m, $n=2.38~\text{mm}^ {-3}$, $c=1500$~m.s$^{-1}$ and $f_0=\omega_0/2\pi = 203$~kHz ($\lambda_0 = 7.4$~mm).
We address both cases of (i) a perfectly random sample and (ii) a disordered collection of pairs oriented along the z-direction and separated by a distance $d=250~\mu$m. We compute the propagation of an axisymmetric Gaussian beam travelling along the z-direction through a thin slab of the metamaterials (see Fig.~\ref{spectre}). We first probe the pressure field at the center of the cloud and obtain the spectra reported in Fig.~\ref{spectre}.
The left plot [Fig.1(a)] shows the simulation results in the fully disordered case. The spectrum exhibits the classical features for such samples, notably resonant modes of the slab at low frequencies and sharp scatterer resonances near $\omega_0$~\cite{lemoult2010resonant,lanoy2015subwavelength}. Above $\omega_0$, the scatterers' response is out of phase with respect to the incident field and the effective compressibility $\chi_\text{eff}$ becomes negative. Because the scattering is completely isotropic ($f(0)=f(\pi)$), the effective density remains unaffected ($\rho_\text{eff}=\rho_0$). As a result, the effective wavevector $k_\text{eff}=\omega\sqrt{\rho_\text{eff}\chi_\text{eff}}$ becomes imaginary and the coherent field is evanescent, so that propagation ceases (band gap). The spectra show that the density of states vanishes beyond $\omega_0$ (colored area) which is consistent with the previous analysis. In the case of the pair-correlated sample (right, [Fig.1(b)]), we observe a narrowing of the band gap together with the appearance of resonant peaks within this band gap. This can be explained considering that the multiple scattering between the two resonators of a single pair may generate a dipolar antisymmetric mode, which can change the sign of $\rho_\text{eff}$. In the region of this dipolar resonance, the real part of $k_\text{eff}$ can then become substantial, thus restoring the propagation.

\begin{figure*}[htb!]
    \centering
      \includegraphics[width=.75\linewidth]{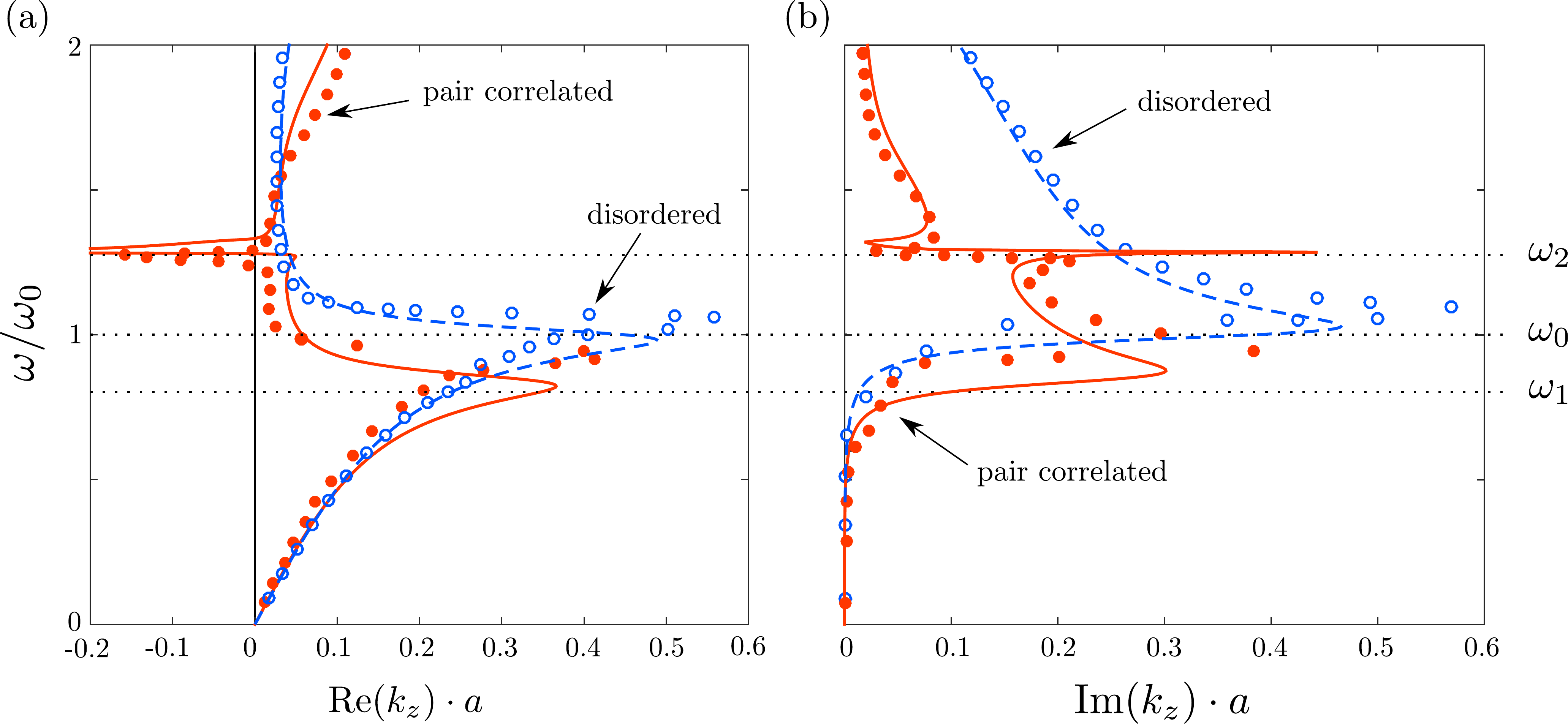}
    \caption{Real (a) and imaginary (b) parts of the reduced wavevector projected along the propagation (Oz) axis for a spatially random sample of $a=100~\mu$m scatterers (blue open symbols) and for a pair correlated disordered sample of the same scatterers (red symbols). The concentration is  $n=2.38~\text{mm}^ {-3}$. Solid lines refer to analytical predictions obtained by applying Eqs. (1) and (2) in the case of an isotropic scatterer (blue dashed line) and in the case of a pair of isotropic scatterers (red solid line). The horizontal dotted lines indicate positions of the single scatterer resonance frequency ($\omega_0$), symmetrical mode frequency ($\omega_1$) and antisymmetrical mode frequency ($\omega_2$).}\label{disp}
\end{figure*}

To go further into the analysis, we directly determine $k_\text{eff}(\omega)$ by simulating the transmission and reflection coefficients of the meta-slab and applying the method detailed in Ref.~\onlinecite{fokin2007method}. Hence we obtain the dispersion relation along the $z$ direction for both the perfectly disordered and pair correlated samples (Fig.~\ref{disp}, symbols).
In the case of the perfectly random sample (blue symbols), one can clearly identify the classical polariton-like evolution for which the wave vector is essentially real below $\omega_0$ and purely imaginary within the band gap~\cite{lagendijk1996polariton,kafesaki2000air,liu2000locally}.
However, the medium populated with pairs of resonators shows a significantly different behavior: (i) the transition from a propagative to an evanescent state occurs at lower frequencies, (ii) the real part of the wave number exhibits a sharp peak while its imaginary part nearly cancels for $\omega/\omega_0\simeq1.3$, thus explaining the opening of a propagating band near this frequency, 
and (iii) this band has negative values of Re($k_z$), consistent with left-handed behavior for the meta-slab in this frequency range. The analytic description (solid lines) was obtained using Eqs. (1) and (2). In the case of the random sample we simply have $f(0)=f(\pi)=f$ and predict behavior (blue dashed line) that is consistent with the numerical results. In order to describe the pair-correlated sample, we took into account the multiple scattering to compute the wave amplitude scattered off a pair of resonators~\cite{lanoydoc,suppl}:
\begin{align}
\frac{f(0)+f(\pi)}{2} & = \frac{2a}{\left( \frac{\omega_0}{\omega}\right)^2 - (1+\frac{a}{d}) - 2\ii  k_0 a} \\
\frac{f(0)-f(\pi)}{2}  & = \frac{k_0^2 d^2 a/2}{\left( \frac{\omega_0}{\omega}\right)^2 - (1-\frac{a}{d})-\text{i} k_0^3 a d^2 /6}.
\end{align}
Note that these expressions show the existence of a monopolar resonance at $\omega_1=\omega_0/(1+a/d)^{1/2}$ and a dipolar resonance at $\omega_2=\omega_0/(1-a/d)^{1/2}$. After substituting Eqs. (3) and (4) in (1) and solving for $k_\text{eff}(\omega)$, we obtained the dispersion relation represented in Fig.~\ref{disp} (red solid line). Although the frequency of the monopolar resonance seems to be slightly underestimated by the model for both kinds of sample, we observe very good agreement in the region of the negative band that is found to open at frequency $\omega_2$.

\begin{figure}[htb!]
    \centering
      \includegraphics[width=.93\linewidth]{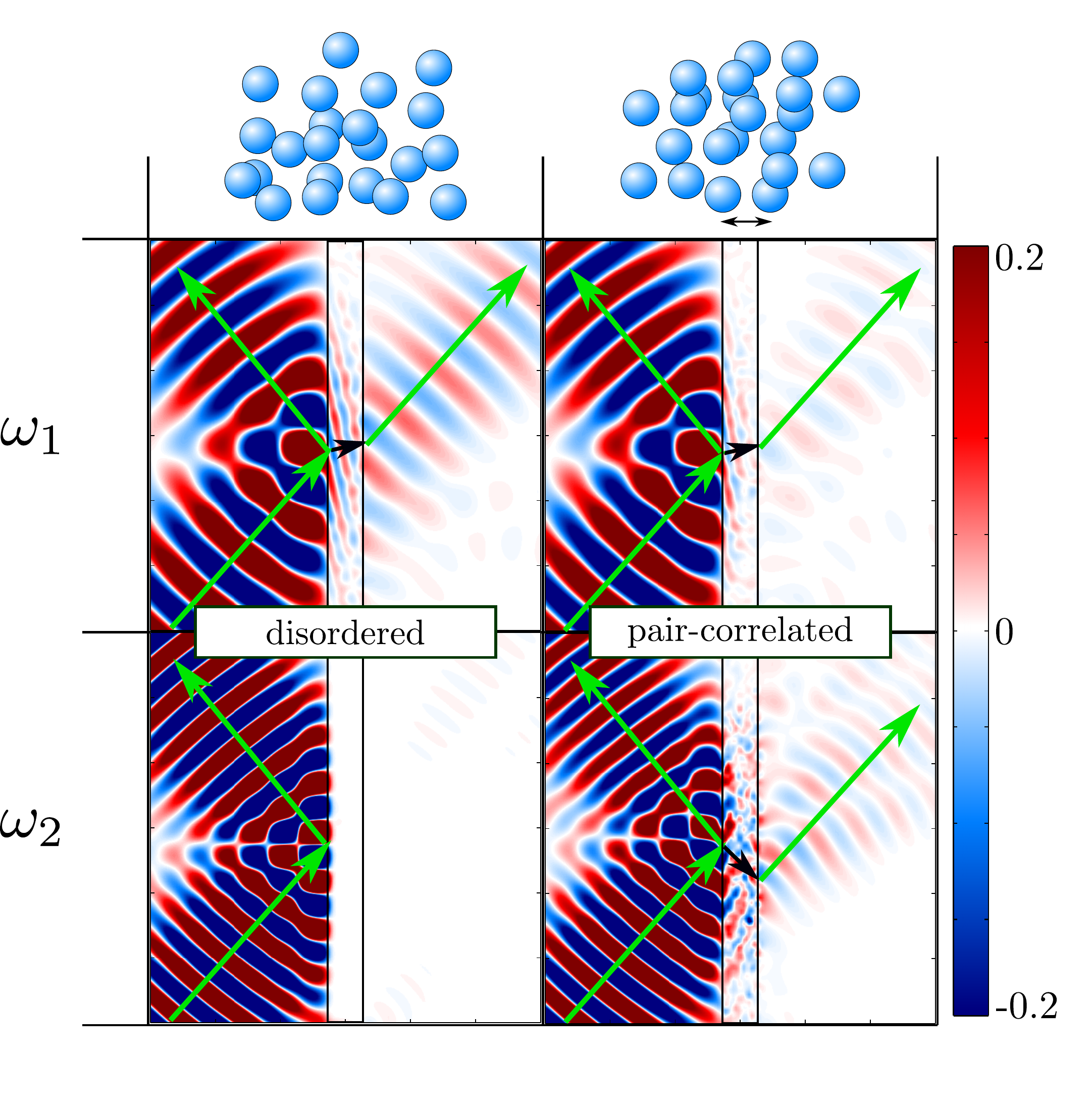}
    \caption{Refraction patterns for an obliquely incident Gaussian beam ($\theta=51.5^{\circ}$). Left: Perfectly random sample. Right: pair-correlated sample. Two frequencies are investigated: $\omega_1$ (top) and $\omega_2$ (bottom). Arrows indicate the propagation directions. The edges of each quadrant are $6~$cm long and the sample thickness is $5$~mm. Parameters: $a=100~\mu$m, $d=250~\mu$m, $n=2.38~\text{mm}^ {-3}$.}\label{refraction}
\end{figure}
As the scatterer pairs are aligned along a specific direction (Oz), the question arises whether the negative density behavior is limited to normal incidence only. In Fig.~\ref{refraction}, we investigate the case of a Gaussian beam impinging on the medium at an angle of $\theta=51.5^{\circ}$. We compare the numerical results for both the random (left) and the pair-correlated (right) samples. We represent the field maps for a monochromatic excitation at frequencies  $\omega_1$ (top) and $\omega_2$ (bottom). Note that the pressure fields corresponding to a single realization of disorder (not shown here) generally feature very sharp hotspots within the sample. It is hence more instructive to compute the coherent field which can be retrieved by averaging the field over $1,000$ statistically independent configurations. As can be anticipated from Fig.~\ref{disp}, both samples show similar refractive behavior at frequency $\omega_1$. At frequency $\omega_2$ however, they exhibit thoroughly different transmissions. Because propagation is forbidden in the perfectly disordered sample (Fig.~\ref{refraction}, bottom left), the incoming field is entirely reflected at the first interface. However, in the pair-correlated case, a substantial part of the incident wave is transmitted at the same frequency (Fig.~\ref{refraction}, bottom right). Note the negative bending of the wavefronts inside the sample, and the downward shift in the position of the transmitted beam, both of which indicate negative refraction for the pair correlated sample at $\omega_2$. Overall, Figs.~\ref{spectre}-\ref{refraction} provide unambiguous evidence that introducing a pair correlation within a disordered sample of monopolar resonators can cause double negativity in a transparency window where negative refraction occurs.

\begin{figure}[htb!]
    \centering
      \includegraphics[width=.95\linewidth]{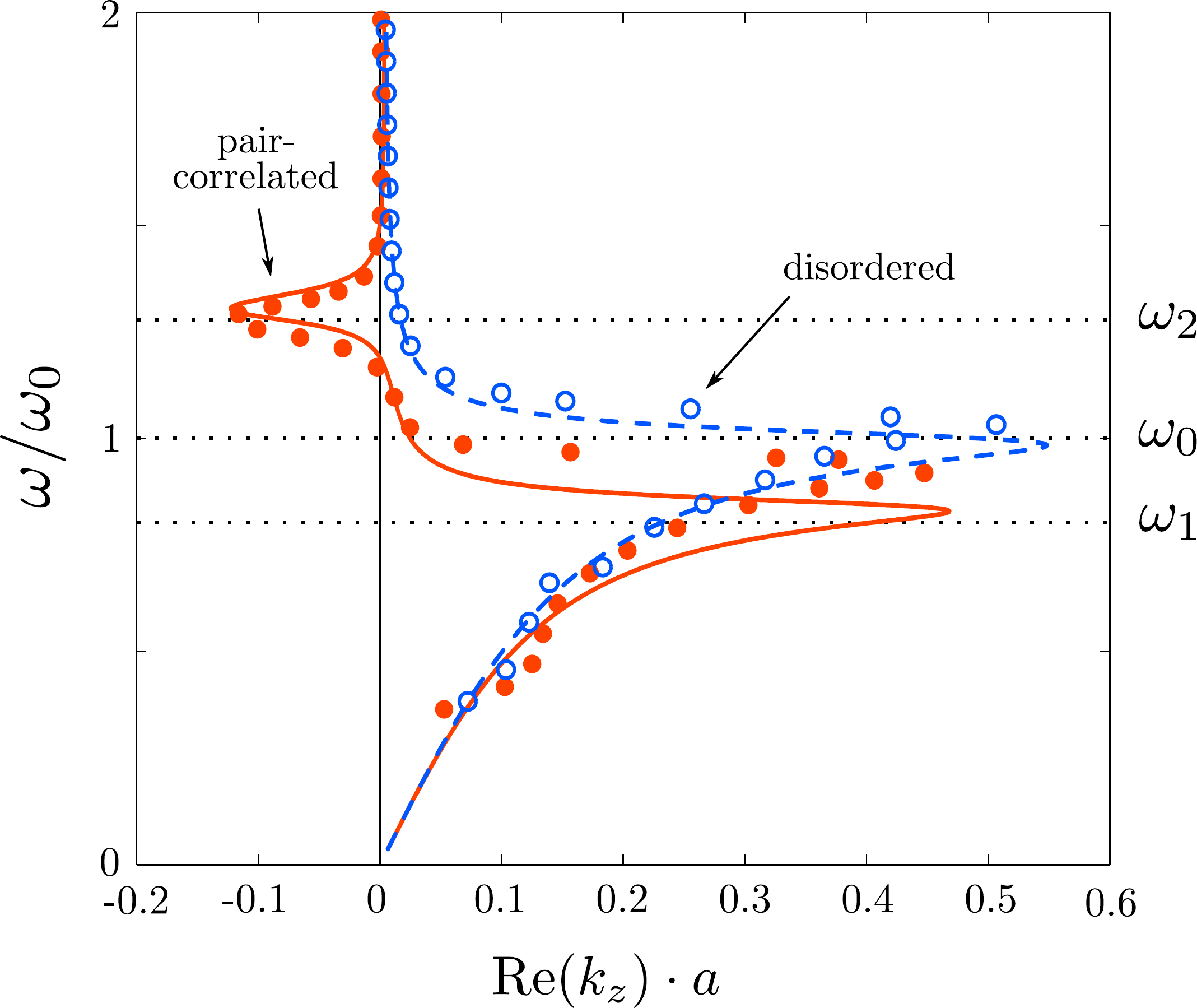}
    \caption{Dispersion curves along the $z$ direction for a random (blue open symbols) and for a pair-correlated disordered sample (red solid symbols) of air bubbles in a yield stress fluid. Dashed and solid lines refer to analytical predictions.
    The horizontal dotted lines indicate frequencies $\omega_0$, $\omega_1$ and $\omega_2$. Parameters: $a=100~\mu$m, $d=250~\mu$m, $n=2.38~\text{mm}^ {-3}$.}\label{dispbub}
\end{figure}

To enable meaningful comparison with possible experiments, losses need to be taken into account. Thanks to our analytical derivation, one can easily establish the following criterion for double negativity~\cite{suppl}:
\begin{equation}
\pi n d^2 a>\delta+k_0^3 a d^2/6.
\label{criterion}
\end{equation}
When $\delta=0$, we simply need to make sure that $n>n_\text{min}=k_0^3/(6\pi)$, which is very easy to satisfy since  $n_\text{min} \simeq 0.07$~mm$^{-3}$ at $\omega_2$. One can also estimate the width of the negative band~\cite{suppl}, yielding here a value of $5\%$, which is consistent with our observations. However, when $\delta \neq 0$,  (\ref{criterion}) might indicate that double negativity no longer occurs anymore. For instance, let us consider a sample of pair-correlated air bubbles with $a=100~\mu$m trapped in a yield stress fluid. In this scenario, the dissipation originates from thermal and viscous processes and one can reasonably consider a resulting damping factor of $\delta=0.05$~\cite{devin1959survey, prosperetti1977thermal}. For such a value, the minimum concentration allowing (\ref{criterion}) to be met amounts to $n_\text{min}=2.70$~mm$^{-3}$, which is slightly higher than the concentration considered above ($n=2.38$~mm$^{-3}$). As a consequence, the doubly negative behavior is not expected anymore. However, the dispersion curve corresponding to this pair-correlated sample (see Fig.~\ref{dispbub}) still exhibits a significant negative band in the region of $\omega_2$, for both numerical and analytical calculations.
Recent works~\cite{brunet2015soft, dubois2014spaces} shed light on this apparent contradiction. In fact, when losses are taken into account, left-handed behavior can be observed even without satisfying the double negativity criterion.
A more accurate criterion to assess for the opening of the negative band at $\omega_2$ is~\cite{suppl}:
\begin{equation}
4\pi n d a^2 >  (\delta + 2k_0a) (\delta+k_0^3 a d^2/6)
\label{criterion2}
\end{equation}
which is easier to satisfy than~(\ref{criterion}).

To conclude, we have demonstrated how to create a 3D disordered double negative metamaterial composed solely of monopolar resonators. Our approach provides a novel pathway to achieving double negativity in a disordered metamaterial, since it is commonly believed that individual monopolar and dipolar resonators are both needed in acoustic systems.  Because the scattering process is relatively easy to describe mathematically when only monopolar resonators are involved, we are able to provide convincing analytic support for our numerical calculations, which are based on the MST. In particular, we show that multiple scattering between two neighboring resonators can introduce a dipolar resonance and thus locally affect the dynamic density of the material. By designing a disordered medium with pair-wise spatial correlations between the monopolar resonators,
we ensure the effectiveness of this pair-wise coupling, which then affects the global properties of the entire material.
We have also established the criteria that must be satisfied for negative refraction to occur in this pair-correlated material, both for the ideal case where dissipation can be neglected and in the more realistic situation when losses are taken into account.  It is interesting to note that in the perfectly ordered case of a two-dimensional crystal of Helmholtz resonators, an analogous subwavelength coupling leading to negative refraction has recently been demonstrated~\cite{kaina2015negative}. More generally, our study illustrates a powerful feature of locally resonant metamaterials for which deeply subwavelength modifications (e.g., in local structure) can induce major changes in behavior. Another remarkable example is the case of ``hyperuniform materials", which can be both dense and transparent due to the microscopic-scale ordering of the particles~\cite{Leseur:16}, and it is reasonable to expect that many other, equally striking, examples are still to be explored.

This work is supported by LABEX WIFI  (Laboratory of Excellence within the French Program ``Investments for the Future'') under references ANR-10-LABX-24 and ANR-10-IDEX-0001-02 PSL*. We thank Direction G\'en\'erale de l'Armement (DGA) for financial support to M.L..  J.H.P. would like to acknowledge support from the NSERC Discovery Grant program.


\begin{thebibliography}{10}


\bibitem{vellekoop2007focusing}
I.~M.~Vellekoop and A.~P.~Mosk,
\newblock {Opt. Lett.} \textbf{32}, 23092311 (2007).

\bibitem{cassereau1992time}
D.~Cassereau and M.~Fink,
\newblock {IEEE Trans. Ultrason. Ferroelectr. Freq. Control} \textbf{39}, 579592 (1992).

\bibitem{tao2008metamaterial}
H.~Tao, N.~I.~Landy, C.~M.~Bingham, X.~Zhang, R.~D.~Averitt, and W.~J.~Padilla,
\newblock {Opt. Express} \textbf{16}, 71817188 (2008).

\bibitem{landy2008perfect}
N.~I.~Landy, S.~Sajuyigbe, J.~J.~Mock, D.~R.~Smith, and W.~J.~Padilla,
\newblock {Phys. Rev. Lett.} \textbf{100}, 207402 (2008).

\bibitem{mei2012dark}
J.~Mei, G.~Ma, M.~Yang, Z.~Yang, W.~Wen, and P.~Sheng,
\newblock {Nat. Commun.} \textbf{3}, 756 (2012).

\bibitem{alu2005achieving}
A.~Al{\`u} and N.~Engheta,
\newblock {Phys. Rev. E} \textbf{72}, 016623 (2005).

\bibitem{leonhardt2006optical}
U.~Leonhardt,
\newblock {Science} \textbf{312}, 17771780 (2006).

\bibitem{pendry2006controlling}
J.~B.~Pendry, D. Schurig, and D. R. Smith,
\newblock {Science} \textbf{312}, 17801782 (2006).

\bibitem{pendry2000negative}
J.~B.~Pendry,
\newblock {Phys. Rev. Lett.} \textbf{85}, 3966 (2000).

\bibitem{shelby2001experimental}
R.~A.~Shelby, D.~R.~Smith, and S.~Schultz,
\newblock {Science} \textbf{292}, 7779 (2001).

\bibitem{smith2000composite}
D.~R.~Smith, W.~J.~Padilla, D.~C.~Vier, S.~C.~Nemat-Nasser, and S.~Schultz,
\newblock {Phys. Rev. Lett.} \textbf{84}, 4184 (2000).

\bibitem{notomi2000theory}
M.~Notomi,
\newblock {Phys. Rev. B} \textbf{62}, 10696 (2000).

\bibitem{mizuguchi2003focusing}
J.~Mizuguchi, Y.~Tanaka, S.~Tamura, and M.~Notomi,
\newblock {Phys. Rev. B} \textbf{67}, 075109 (2003).

\bibitem{foteinopoulou2003refraction}
S.~Foteinopoulou, E.~N.~Economou, and C.~M.~Soukoulis,
\newblock {Phys. Rev. Lett.} \textbf{90}, 107402 (2003).

\bibitem{cubukcu2003electromagnetic}
E.~Cubukcu, K.~Aydin, E.~Ozbay, S.~Foteinopoulou, and
  C.~M.~Soukoulis,
\newblock {Nature} \textbf{423}, 604605 (2003).

\bibitem{parimi2003photonic}
P.~V.~Parimi, W.~T.~Lu, P.~Vodo, and S.~Sridhar,
\newblock {Nature} \textbf{426}, 404404 (2003).

\bibitem{yang2004focusing}
S.~Yang, J.~H.~Page, Z.~Liu, M.~L.~Cowan, C.~T.~Chan, and P.~Sheng,
\newblock {Phys. Rev. Lett.} \textbf{93}, 024301 (2004).

\bibitem{zhang2004negative}
X.~Zhang and Z.~Liu,
\newblock {Appl. Phys. Lett.} \textbf{85}, 341343 (2004).

\bibitem{sukhovich2008negative}
A.~Sukhovich, L.~Jing, and J.~H.~Page,
\newblock {Phys. Rev. B} \textbf{77}, 014301 (2008).

\bibitem{sukhovich2009experimental}
A.~Sukhovich, B.~Merheb, K.~Muralidharan, J.~O.~Vasseur, Y.~Pennec, P.~A.~Deymier, and J.~H.~Page,
\newblock {Phys. Rev. Lett.} \textbf{102}, 154301 (2009).

\bibitem{foldy}
L.~L.~Foldy,
\newblock {Phys. Rev.} \textbf{67}, 107 (1945).

\bibitem{waterman1961multiple}
P.~C.~Waterman and R.~Truell,
\newblock {J. Math. Phys.} \textbf{2}, 512537 (1961).

\bibitem{ding2007metamaterial}
Y.~Ding, Z.~Liu, C.~Qiu, and J.~Shi,
\newblock {Phys. Rev. Lett.} \textbf{99}, 093904 (2007).

\bibitem{lee2010composite}
S.~H.~Lee, C.~M.~Park, Y.~M.~Seo, Z.~G.~Wang, and C.~K.~Kim,
\newblock {Phys. Rev. Lett.} \textbf{104}, 054301 (2010).

\bibitem{li2004double}
J.~Li and C.~T.~Chan,
\newblock {Phys. Rev. E} \textbf{70}, 055602 (2004).


\bibitem{brunet2015soft}
T.~Brunet, A.~Merlin, B.~Mascaro, K.~Zimny, J.~Leng,
  O.~Poncelet, C.~Arist{\'e}gui, and O.~Mondain-Monval,
\newblock {Nat. Mater.} \textbf{14}, 384388 (2015).

\bibitem{mie1908beitrage}
G.~Mie,
\newblock {Ann. Phys. (Berl.)} \textbf{330}, 377445 (1908).

\bibitem{lax1952multiple}
M.~Lax.
\newblock {Phys. Rev.} \textbf{85}, 621 (1952).

\bibitem{lanoy2015subwavelength}
M.~Lanoy, R.~Pierrat, F.~Lemoult, M.~Fink, V.~Leroy, and A.~Tourin,
\newblock {Phys. Rev. B} \textbf{91}, 224202 (2015).

\bibitem{lemoult2010resonant}
F.~Lemoult, G.~Lerosey, J.~de~Rosny, and M.~Fink,
\newblock {Phys. Rev. Lett.} \textbf{104}, 203901 (2010).

\bibitem{fokin2007method}
V.~Fokin, M.~Ambati, C.~Sun, and X.~Zhang,
\newblock {Phys. Rev. B} \textbf{76}, 144302 (2007).

\bibitem{lagendijk1996polariton}
A.~Lagendijk and B.~A.~van~Tiggelen,
\newblock {Phys. Rep.} \textbf{270}, 143 (1996).

\bibitem{kafesaki2000air}
M.~Kafesaki, R.~S.~Penciu, and E.~N.~Economou,
\newblock {Phys. Rev. Lett.} \textbf{84}, 6050 (2000).

\bibitem{liu2000locally}
Z.~Liu, X.~Zhang, Y.~Mao, Y.~Y.~Zhu, Z.~Yang, C.~T.~Chan, and P.~Sheng,
\newblock {Science} \textbf{289}, 17341736 (2000).

\bibitem{suppl}
For a detailed derivation, see the Supplemental Material.

\bibitem{lanoydoc}
M.~Lanoy, Doctoral Thesis, Universit\'e Paris Denis Diderot (2016).

\bibitem{devin1959survey}
C.~Devin~Jr,
\newblock {J. Acoust. Soc. Am.} \textbf{31}, 6541667 (1959).

\bibitem{prosperetti1977thermal}
A.~Prosperetti,
\newblock {J. Acoust. Soc. Am.} \textbf{61}, 1727 (1977).

\bibitem{dubois2014spaces}
J.~Dubois, C.~Arist{\'e}gui, and O.~Poncelet,
\newblock {J. Appl. Phys} \textbf{115}, 024902 (2014).

\bibitem{kaina2015negative}
N.~Kaina, F.~Lemoult, M.~Fink, and G.~Lerosey,
\newblock {Nature} \textbf{525}, 7781 (2015).

\bibitem{Leseur:16}
O.~Leseur, R.~Pierrat, and R.~Carminati,
\newblock {Optica} \textbf{3}, 763767 (2016).


\end{thebibliography}
\end{document}